\documentclass[psfig]{kapproc} 

\RequirePackage{graphicx}%
\RequirePackage{epsf}%
\RequirePackage{psfig}%

\upperandlowercase
\let\footnote\savefootnote
\let\footnotetext\savefootnotetext 
\let\footnoterule\savefootnoterule 
\kluwerbib 

\begin{document}

\articletitle{Lambda Orionis: A 0.02--50 M$_\odot$ IMF}

\author{D. Barrado y Navascu\'es\altaffilmark{1}, J.R. Stauffer\altaffilmark{2}, 
        J. Bouvier\altaffilmark{3}}

\affil{\altaffilmark{1}LAEFF-INTA (ESA tracking Station), Madrid, SPAIN, \\ 
       \altaffilmark{2}IPAC, Caltech, USA, \\
       \altaffilmark{3}Observatoire de Grenoble, FRANCE}

\email{barrado@laeff.esa.es, stauffer@ipac.caltech.edu, Jerome.Bouvier@obs.ujf-grenoble.fr}

\chaptitlerunninghead{The IMF of Lambda Orionis cluster}


 \begin{abstract}
We derive the initial mass function for the Col~69 
cluster ($~$5 Myr), covering several orders of magnitude in mass
(50 -- 0.02 M$_\odot$).
 \end{abstract}


\section{The Lambda Orionis Star Forming Region}

One of the most amazing areas in the northern, winter sky is
Orion and its myriads of star forming regions (SFR). One of the most 
prominent, albeit not very well studied, is the SFR dominated by the
O8 III star $\lambda$$^1$ Ori, the Lambda Orionis SFR (LOSFR).
Among other structures, it includes   a CO and a dust ring whose
 diameter is about nine deg, the S264 H{\sc II} region,
a large number of IRAS sources, 
the Barnard 30 and 35 dark clouds and a cluster associated with  the
central star, the Lambda Ori cluster (Collinder 69).
As a laboratory for the star formation process (or processes), 
the LOSFR represents a unique environment due to its diversity,
its proximity (400 pc), its age range ($\sim$5 Myr) and low extinction
in the internal area of the ring (A$_V$=0.37).
Prior to our study, several groups have published
studies focused on different aspects of the SFR,
such as the initial discovery by Wade (1957),  the photometric properties of the 
high mass members by Murdin \& Penston (1977; hereafter M\&P77),
a H$\alpha$ survey by Duerr, Imhoff \& Lada (1982),
the analysis of the IRAS data (Zhang et al. 1989) and the 
photometric and spectroscopic search by Dolan \& Mathieu (1999, 2001, 2002; hereafter D\&M).
This last set of studies reached I$_C$=14.5, about 0.5 M$_\odot$. Our 
aim is to carry out a comprehensive study in the LOSFR, reaching well below
the hydrogen burning limit 
(HBL, the substellar border line at $\sim$0.072  M$_\odot$,
as derived by the Lyon models,  Baraffe et al. 1998). Here, we present some initial results
mainly concerning the central area of this fascinating SFR.

\begin{figure}
\sidebyside
{\centerline{\psfig{file=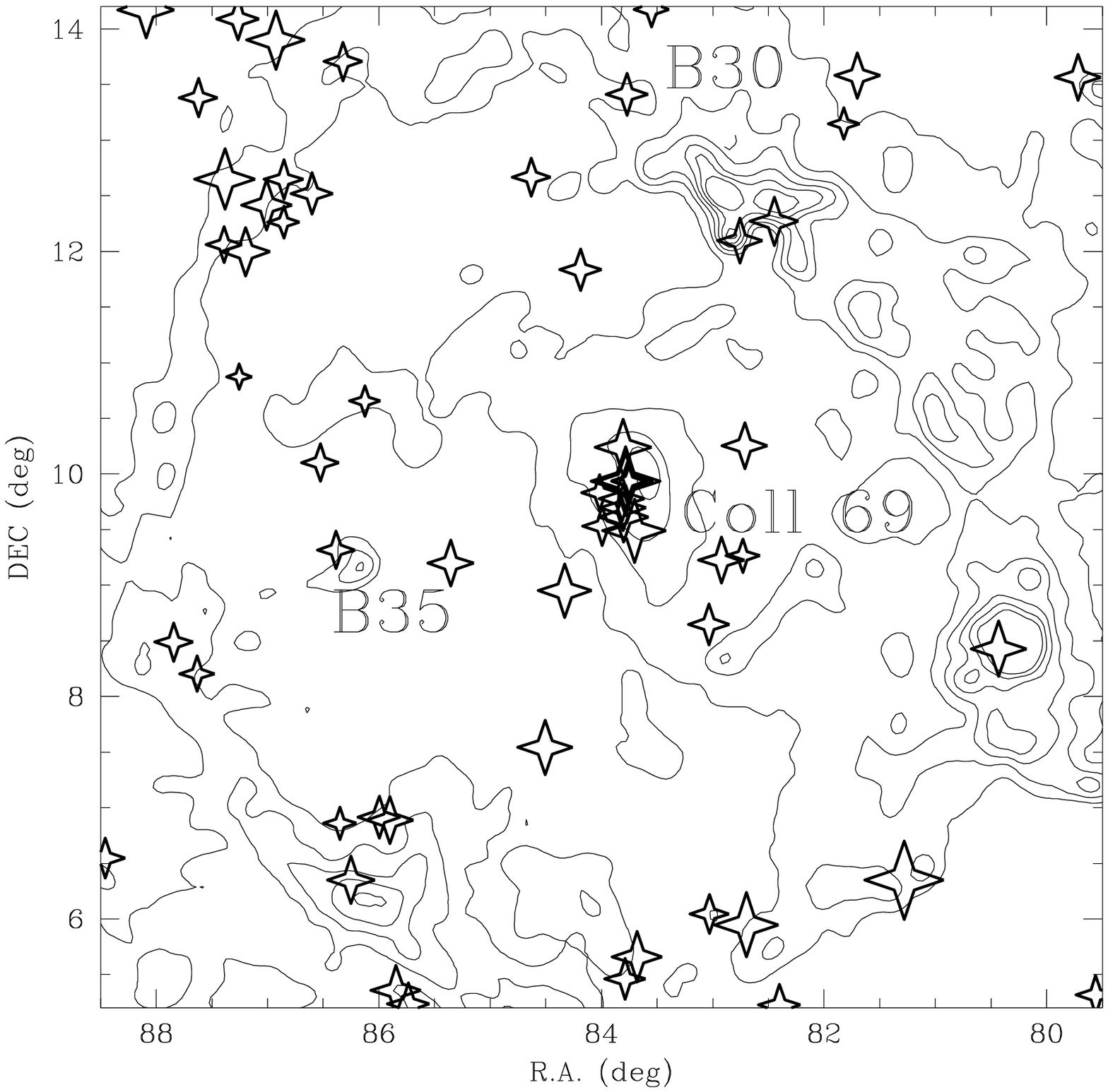,height=2.5in}}
\caption{An IRAS image at 100 $\mu$, showing the LOSFR.}}
{\centerline{\psfig{file=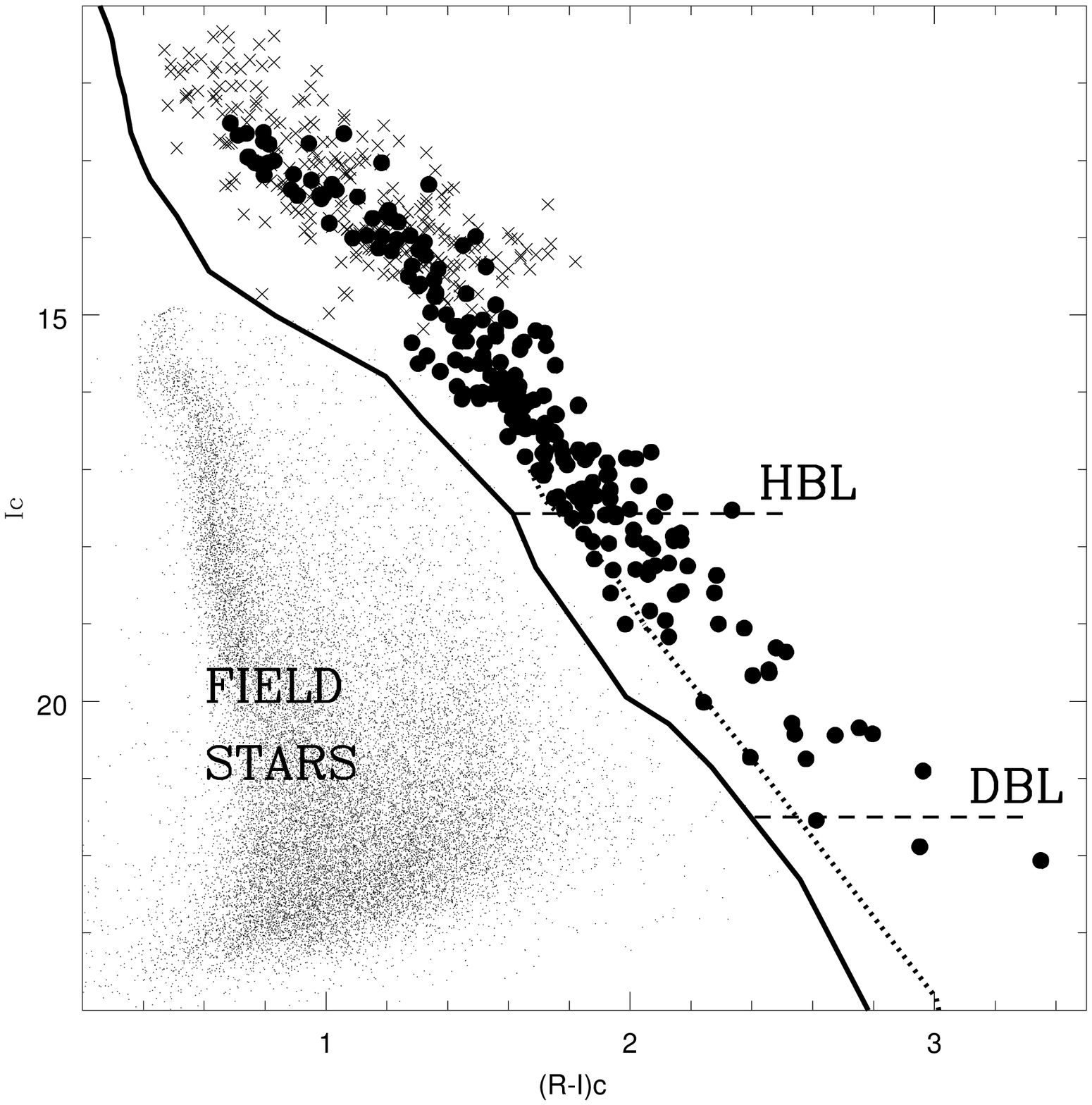,height=2.5in}}
\caption{A deep Color-magnitude diagram in the core of the LOSFR.}}
\end{figure}

\section{The Lambda Orionis cluster, Collinder 69}

The star cluster associated with the star $\lambda$ Orionis
is clearly identified in Figure 1 by the overdensity of B stars
 (four point stars).
In order to identify low mass members of this association, we have
conducted several photometric and spectroscopic campaigns,
 both in the optical and the near infrared. Details regarding 
the first of them,
an optical search --Figure 2-- with the CFHT and the 12K mosaic, supplemented
with 2MASS JHK data 
and spectroscopy from Keck and Magellan, can be found
in Barrado y Navascu\'es et al. (2004, Paper I).
Note that for a 5 Myr age, and based on NextGen and Cond 
(Chabrier et al. 2000) models, 
the hydrogen and deuterium burning limits  are located at I$_C$=17.6
and I$_C$=22, i.e.,  we have sampled the complete brown dwarf domain.

\begin{figure}
\centerline{\psfig{file=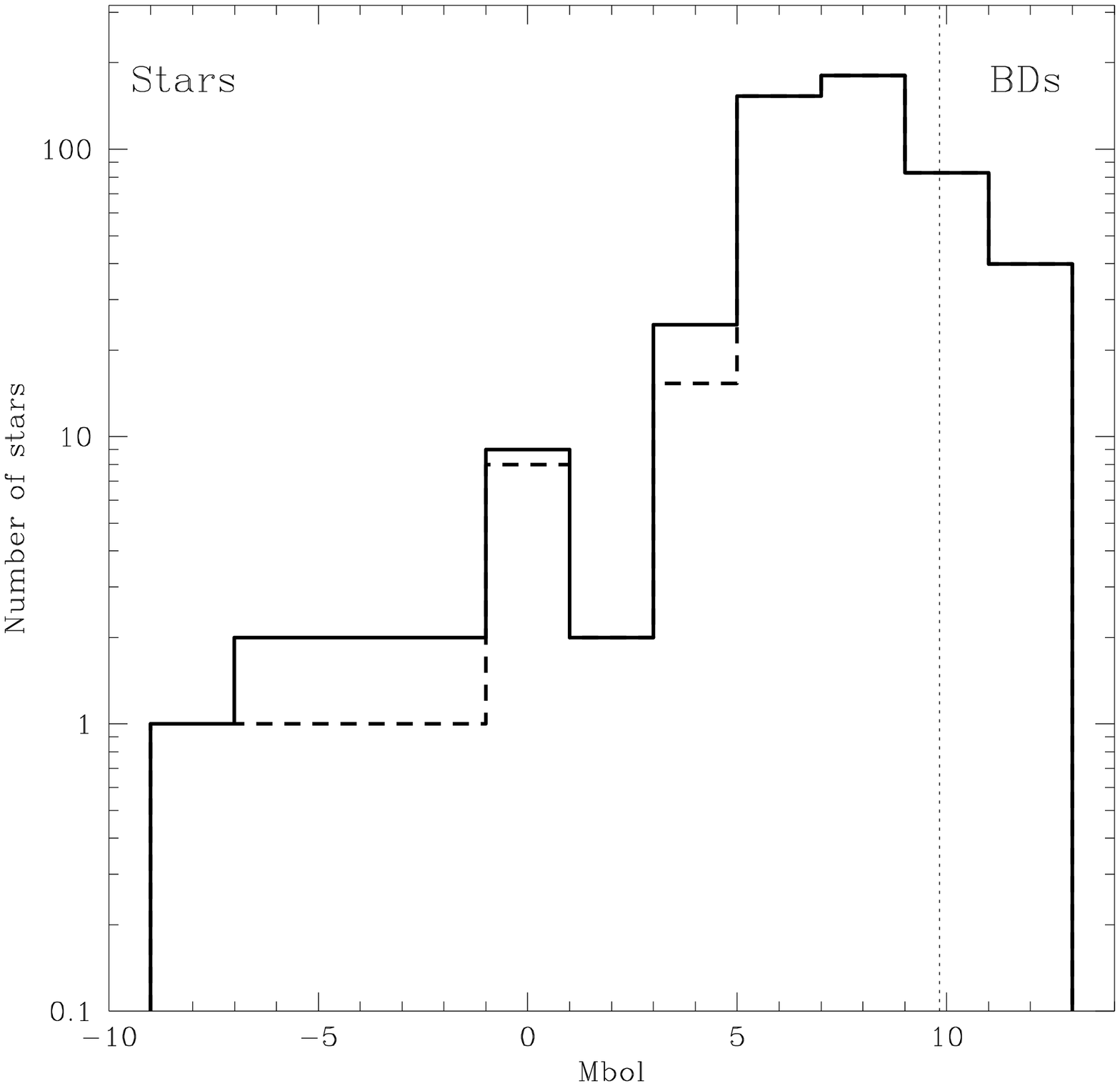,height=2.5in} \psfig{file=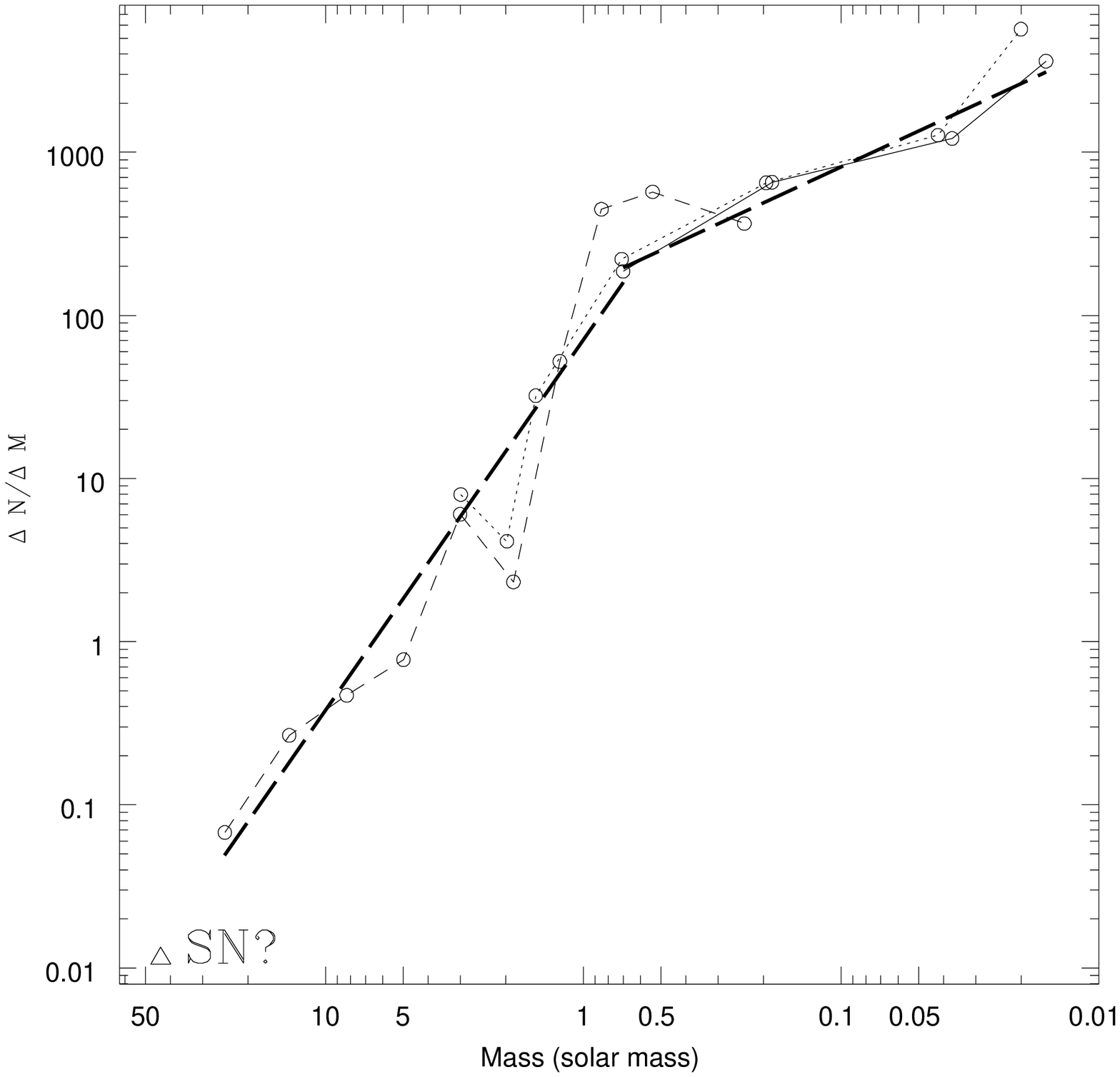,height=2.5in}}
\caption{The Coll~69 Luminosity and Mass Functions.}
\end{figure}
\inxx{captions,figure}

Based on this wealth of data, we have found 170 candidate
members in the range I$_C$=12.5--22.0, which translate
to 1.2--0.01 M$_\odot$ for bona fide members.
Of these 170, 33 have been spectroscopically  observe at low-resolution
(I$_C$=15.2--20.7, M4.5--M8.5 spectral types),
whereas another 25 have medium resolution spectra 
(I$_C$=13.7--17.6, M4--M6.0).
The analysis of the  optical-IR information --see color-magnitude and 
color-color diagrams in Paper I-- indicates that
the pollution rate should be about  30\%.
A similar value is derived from the spectroscopy.

We have also studied other properties of the low mass stellar
and substellar population of the Lambda Orionis cluster --Coll~69--
and compare those properties with previous results from the LOSFR and 
other young associations.  In particular, we have measured the
strength of the lithium doublet at 6708 \AA{} (see Palla \&
Randich, this volume) and H$\alpha$,
another signpost of youth and accretion.  Our targets show lithium
in their medium-resolution spectra, clearly indicating that they are PMS
objects.  On the other hand, by comparison with the criterion for
H$\alpha$ defined in Barrado y Navascu\'es \& Mart\'{\i}n (2003),
we find that Coll~69 has a paucity of CTT stars compared with the
clusters associated with the dark clouds Barnard 30 and 35 (as
already noted by D\&M).  This might be related to the obvious
richness of hot, high mass stars in the first association, compared
with the other two, as is easily appreciated in Figure 1.  D\&M
suggested a scenario to explain these differences, 
with the episodes of star formation in B30 and B35 
(and eventually in the
ring in the LOSFR) triggered by a supernova in Coll~69.  This
catastrophic event is speculated to have removed most of the
disk material around the young stars in Coll~69.  If this scenario
is true, Coll~69 should be older than the other two clusters, and
have a different history of star formation - possibly leading
to a different IMF.

\subsection{The IMF of the Lambda Orionis cluster}

We have derived a Luminosity and Initial Mass Functions (LF and IMF) for the 
bona fide members of Coll~69. 
We emphasize that  our IMF characterize only the 
central part of the cluster.
For more massive stars, we used data from D\&M 
and  M\&P77, scaling the areas in the appropriate way.
Regarding the LF (Figure 3a),
 the solid and dashed lines were computed including
all possible members (solid line)
 and removing the  photometric binaries 
(possible members, dashed line),  respectively.

The IMF is displayed in Figure 3b. 
In order to extend the IMF to the high mass domain, 
we have included data from M\&P97. We have faced  three 
important problems: the in-homogeneity of the photometric data
 (we have derived bolometric magnitudes from $V$ and $Ic$),
  the paucity of massive stars --i.e., the validity of the
LF in that range-- and the selecting of a theoretical models. 
We have used models by Girardi et al. (2002),
D'Antona \& Mazzitelli (1997) and Baraffe et al. (1998), for massive, 
intermediate and low mass objects (dashed, dotted and solid lines), 
 respectively. 
The dip at about 2 M$_\odot$ is due to incompleteness in the sample.
The last point (open triangle) at the massive end correspond to the 
possible SN.  Note that the binning in the 
LF is arbitrary, although the results do not change much by selecting other
values. The slopes of the IMF (bold, dashed segments) are
$\alpha$=2.27$\pm0.08$,  and $\alpha$=0.73$\pm$0.05, for the mass ranges
25--0.70 and 0.70-0.02 M$_\odot$, respectively. 
The figure indicates that the SN hypothesis is compatible with the
 IMF of the cluster.
To the best of our knowledge, this IMF is one of the most complete and 
accurate --due to the reduced internal reddening of the cluster-- 
IMF published in the literature, covering more than three orders of
 magnitude in mass.

\begin{acknowledgments}
DByN is indebted to the Spanish ``Programa Ram\'on y Cajal'', 
PNAyA AYA2001-1124-C02 and PNAyA AYA2003-05355. 
\end{acknowledgments}


\begin{chapthebibliography}{1}
\bibitem[1998]{baraffe98} 
Baraffe I., Chabrier G., Allard F., Hauschildt P. H.,
 1998, A\&A, 337, 403

\bibitem[2002]{barrado2002}
Barrado y Navascu\'es  D.,  et al.
2004, ApJ 610, 1064

 \bibitem[2003]{barradomartin2003} 
Barrado y Navascu\'es D., Mart\'{\i}n E.L.,
2003, AJ 126, 2997

\bibitem[2000]{Chabrier2000}
Chabrier G.,  Baraffe I., , Allard F., 
Hauschildt P., 
2000, ApJ,  542, L119.

\bibitem[1997]{DAntona1997}
D'Antona, F., \& Mazzitelli, I., 1997, Mem.S.A.It. 68, 807

\bibitem[]{}
Dolan C.J \& Mathieu R.D.,
1999, AJ 118, 2409

\bibitem[]{}
Dolan C.J \& Mathieu R.D.,
2001, AJ 121, 2124

\bibitem[]{}
Dolan C.J \& Mathieu R.D.,
2002, AJ 123, 387

\bibitem[]{}
Duerr R., Imhoff C.L.,  Lada C.J.,
1982, ApJ 261, 135

\bibitem[]{}
Girardi L.,  et al. 
2002, A\&A 391, 195

\bibitem[1997]{murdin1997}
Murdin P., \& Penston M.V.,
1997, MNRAS 181, 657

\bibitem[]{}
Wade C.M.,
1957, AJ 62, 148

\bibitem[]{}
Zhang C.Y., et al.
1989, A\&A 218, 231


\end{chapthebibliography}
\end{document}